\documentclass[preprint,aps,12pt,showpacs,nofootinbib,tightenlines]{revtex4}
\usepackage{amsmath}
\usepackage{amssymb}
\usepackage{epsfig}
\usepackage{graphicx}
\begin{document}
\preprint{KEK TH-1101}
\newcommand{\beq}{\begin{eqnarray}}
\newcommand{\eeq}{\end{eqnarray}}

\newcommand{\bxsga}{B\to X_s \gamma}
\newcommand{\brbxsga}{{\cal B}(B\to X_s \gamma)}
\newcommand{\bzbzb}{ B_d^0 - \hat{B}_d^0 }

\newcommand{\bsga}{  b\to s \gamma}
\newcommand{\bdga}{  b\to d \gamma}
\newcommand{\bvga}{  B\to V \gamma }
\newcommand{\bksga}{ B\to K^* \gamma}
\newcommand{\brhoga}{B\to \rho \gamma}

\newcommand{\brbkz}{{\cal B}(B\to \overline{K}^{*0} \gamma)}
\newcommand{\brbkm}{{\cal B}(B\to K^{*-} \gamma)}
\newcommand{\brbrm}{{\cal B}(B\to \rho^- \gamma)}
\newcommand{\brbrz}{{\cal B}(B\to \rho^0 \gamma)}

\newcommand{\calb}{ {\cal B}}
\newcommand{\acp}{ {\cal A}_{CP}}
\newcommand{\oas}{ {\cal O} (\alpha_s)}

\newcommand{\mt}{m_t}
\newcommand{\mw}{M_W}
\newcommand{\mhp}{M_{H}}
\newcommand{\muw}{\mu_W}
\newcommand{\mub}{\mu_b}
\newcommand{\dmd}{\Delta M_{B_d} }
\newcommand{\ltt}{\lambda_{tt} }
\newcommand{\lbb}{\lambda_{bb} }
\newcommand{\rhob}{\hat{\rho} }
\newcommand{\etab}{\hat{\eta} }

\newcommand{\smallsm}{{\scriptscriptstyle SM}}
\newcommand{\smallyy}{{\scriptscriptstyle YY}}
\newcommand{\smallxy}{{\scriptscriptstyle XY}}
\newcommand{\smallnp}{{\scriptscriptstyle NP}}

\newcommand{\tab}[1]{Table \ref{#1}}
\newcommand{\fig}[1]{Fig.\ref{#1}}
\newcommand{\real}{{\rm Re}\,}
\newcommand{\im}{{\rm Im}\,}
\newcommand{\non}{\nonumber\\ }

\title{Implications of new data in charmless B decays}
\author{ Yue-Liang Wu$^a$}
\email{ylwu@itp.ac.cn} 
%
\author{Yu-Feng Zhou$^b$ }
\email{zhou@post.kek.jp}
\author{Ci Zhuang$^a$ }
\email{zhuangc@itp.ac.cn} \affiliation{(a)\ Kavli Institute for
Theoretical Physics China (KITPC) at the Chinese Academy of Sciences
\\ Institute of theoretical physics, Chinese Academy of Science, Beijing 100080, China \\
(b)\ Theory Group, KEK, Tsukuba 305-0801, Japan}
\date{\today}

\begin{abstract}
Based on the latest experimental data of $B \to \pi\pi$ and $\pi K$
modes, a model-independent analytical analysis is presented. The
CP-averaged branching ratio difference $\Delta R = R_c - R_n$ in
$B\to \pi K$ decays with $R_c = 2Br(\pi^0K^-)/Br(\pi^-\bar{K}^0)$
and $R_n =Br(\pi^+K^-)/2Br(\pi^0\bar{K}^0)$ is reduced though it
remains larger than the prediction from the standard model(SM) as
both measured $R_n$ and $R_c$ are enhanced, which indicates that a
room for new physics becomes smaller. The present data of $\pi\pi$
decay reduce the ratio $|C/T|$ from the previous value of
$|C/T|\simeq 0.8 $ to $|C/T| \simeq 0.65$, which is still larger
than the theoretical estimations based on QCD factorization and
pQCD.  Within SM and flavor SU(3) symmetry, the current $\pi K$ data
also diminish the ratio $|C'/T'|$ from the previous value $|C'/T'|
\simeq 2$ to $|C'/T'| \simeq 1.16$ with a large strong phase
$\delta_{C'} \simeq -2.65$, while its value remains much larger than
the one extracted from the $\pi \pi$ modes. The direct CP violation
$A_{CP}(\pi^0\bar{K}^0)$ is predicted to be $A_{CP}(\pi^0\bar{K}^0)
= -0.15\pm0.03$, which is consistent with the present data. Two
kinds of new effects in both strong and weak phases of the
electroweak penguin diagram are considered. It is found that both
cases can reduce the ratio to  $|C'/T'| = 0.40\sim 0.80$ and lead to
roughly the same predictions for CP violation in $\pi^0 K^0$.
\end{abstract}


\maketitle

\newpage

\section{Introduction}

The measurements of hadronic charmless $B$ decays at the two
$B$-factories become more and more accurate. Currently, all the
branching ratios of $B \rightarrow \pi \pi$ and $\pi K$ modes have
been measured with good accuracy\cite{HFAG,ichep06} and a large direct
CP violation has well been established in $\pi^+ K^-$ mode
\cite{BB1,BB2}.
Recently, the BaBar and Belle collaborations have reported their
updated results which show a better agreement with the Standard
Model\cite{ichep06}. The new world average is summarized in Table.I.
In $\pi\pi$ and $\pi K$ decay modes, one can define the following
ratios:
 \beq
R_{00} &=& \frac{2Br(\pi^0\pi^0)}{Br(\pi^+\pi^-)},
\eeq
and
\beq
R_c &=& \frac{2Br(\pi^0K^-)}{Br(\pi^-\bar{K}^0)},\qquad
R_n = \frac{Br(\pi^+K^-)}{2Br(\pi^0\bar{K}^0)}.
\eeq

A relatively  large $R_{00}$ and $R_n$ or $R_c$ deviating significantly
from unity are usually referred to $\pi\pi$ and $\pi K$ puzzles
respectively. Their implications have been investigated by many
groups\cite{BF,MY,NK,WZ1,WZ2,LMS,GR,CL,HM,CG,BH,IDL,LM,SB,WZZ}.
Compared to the old data which give $R_{00} = 0.67\pm0.14 , R_n =
0.82\pm0.08 $ and $R_c = 1.00\pm0.06 $, the latest data indicate a
reduction of $R_{00}$ and enhancement of $R_n$ and $R_c$, i.e.,
$R_{00} = 0.50\pm0.08 , R_n = 1.00\pm0.07$ and $R_c = 1.11\pm0.08 $.
Namely $R_{00}$ and $R_n$ are moving closer to the SM estimation. On
the contrary, $R_c$ deviates from unity now. The ratio difference
$\Delta R = R_c-R_n$ is reduced in comparison with the previous
result, but it is still puzzling since the ratio difference in the
SM is of the order $O(|P_{EW}/P|^2) \simeq O(10^{-2})$. Within the
SM, both the $\pi\pi$ and $\pi K$ puzzles can be accounted for by
large color suppressed tree diagrams $C$ and $C'$. However, the
previous analysis showed that $|C/T|\simeq 0.7$ in $\pi\pi$ decays,
while in $\pi K$ decays, $|C'/T'| \simeq 2.0$. Such a large
$|C'/T'|$ may either indicate a breakdown of SU(3) symmetry or new
physics from electroweak penguin sector\cite{WZZ}.

In light of the latest data, it is interesting to make an updated
analysis within the framework of quark flavor topology. Our results
show that in both $\pi\pi$ and $\pi K$ modes, the values of $|C/T|$
and $|C'/T'|$ are reduced and closer to the theoretical
estimations. Numerically, it is found that $|C/T| \simeq 0.65$ and
$|C'/T'| \simeq 1.16$ respectively. We also make predictions for the
direct CP asymmetry and mixing induced CP asymmetry of $B \to
\pi^0\bar{K}^0$ and compare them with the preliminary data of Babar
and Belle. In our previous paper\cite{WZ1}, it has been shown that the
weak phase $\gamma$ can well be determined to be remarkably consistent
with the global standard model fit, which gives $\gamma =
1.08^{+0.17}_{-0.21}$ \cite{CKMfitter}. In the present paper, we shall take the CKM phase
$\gamma$ as an input parameter in analytical analysis.

\section{Analysis with diagrammatic decomposition}

\begin{table}[htb]

\label{data}
\begin{ruledtabular}
\begin{tabular}{llll}
  modes & $B r$($\times 10^{- 6}$) & $A_{CP}$ & $S$\\
  \hline
  $\pi^+ \pi^-$ & $5.2 \pm 0.2$ & $0.39 \pm 0.07$ & $- 0.58 \pm 0.09$\\
  \hline
  $\pi^0 \pi^0$ & $1.31 \pm 0.21$ & $0.36^{+0.33}_{-0.31}$ & \\
  \hline
  $\pi^- \pi^0$ & $5.7 \pm 0.40$ & $0.04 \pm 0.05$ & \\
  \hline
  $\pi^+ K^-$ & $19.7 \pm 0.6$ & $- 0.093 \pm 0.015$ & \\
  \hline
  $\pi^0 \bar{K}^0$($K_S$) & $10.0 \pm 0.6$ & $- 0.12 \pm 0.11$ & $( +
  0.33
  \pm 0.21 )$\\
  \hline
  $\pi^- \bar{K}^0$ & $23.1 \pm 1.0$ & $ 0.009 \pm 0.025$ & \\
  \hline
  $\pi^0 K^-$ & $12.8 \pm 0.6$ & $0.047 \pm 0.026$ & \\
\end{tabular}
\end{ruledtabular}
\caption{ The latest world average data of Charmless B
decays\cite{ichep06,HFAG}}.
\end{table}

The general diagrammatic decomposition of the decay amplitudes for
$B \to \pi\pi(\pi K)$ decays can be expressed as (see, e.g. \cite{ISO}):
\begin{eqnarray}
  \bar{\mathcal{A}} ( \pi^+ \pi^- ) & = & - \left[ \lambda_u ( T + E
  -P-P_A - \frac{2}{3} P^C_{EW} ) -
  \lambda_c (P+P_A + \frac{2}{3} P^C_{EW}
  ) \right] , \nonumber\\
  \bar{\mathcal{A}} ( \pi^0 \pi^0 ) & = & - \frac{1}{\sqrt{2}} \left[
  \lambda_u ( C - E + P + P_A - P_{EW} - \frac{1}{3} P^C_{EW} )
  - \lambda_c ( - P - P_A + P_{EW} + \frac{1}{3} P^C_{EW} )
  \right] , \nonumber\\
  \bar{\mathcal{A}} ( \pi^0 \pi^- ) & = & - \frac{1}{\sqrt{2}} \left[
  \lambda_u ( T + C - P_{EW} - P^C_{EW} ) - \lambda_c (
  P_{EW} + P^C_{EW} ) \right] ,
\end{eqnarray}
with $\lambda_u = V_{ub}V_{ud}^* =
A\lambda^3(\rho-i\eta)(1-\lambda^2/2), \lambda_c = V_{cb}V_{cd}^*
= -A\lambda^3$ and:
\begin{eqnarray}
  \bar{\mathcal{A}} ( \pi^+ K^- ) & = & - \left[ \lambda_u^s ( T' - P' -
  \frac{2}{3} P^{C'}_{EW} ) - \lambda_c^s ( P' + \frac{2}{3}
  P^{C'}_{EW} ) \right] ,
\nonumber\\
  \bar{\mathcal{A}} ( \pi^0 \overline{K^{}}^0 ) & = & - \frac{1}{\sqrt{2}}
  \left[ \lambda_u^s ( C' + P' - P'_{EW} - \frac{1}{3} P^{C'}_{EW}
  ) - \lambda_c^s ( - P' + P'_{EW} + \frac{1}{3} P^{C'}_{EW} )
  \right] ,
  \nonumber\\
  \bar{\mathcal{A}} ( \pi^- \overline{K^{}}^0 ) & = & \lambda_u^s ( A' - P' +
  \frac{1}{3} P^{C'}_{EW} ) - \lambda_c^s ( P' - \frac{1}{3}
  P^{C'}_{EW} ) ,
  \\
  \bar{\mathcal{A}} ( \pi^0 K^- ) & = & - \frac{1}{\sqrt{2}} \left[
  \lambda_u^s ( T' + C' + A' - P' - P'_{EW} - \frac{2}{3} P^{C'}_{EW}
  ) - \lambda_c^s ( P' + P'_{EW} + \frac{2}{3} P^{C'}_{EW} )
  \right] , \nonumber
\end{eqnarray}
with $\lambda_u^s = V_{ub} V_{us}^{\ast} = A \lambda^4 ( \rho - i
\eta )$, and $\lambda_c^s = V_{cb} V_{cs}^{\ast} = A \lambda^2 ( 1 -
\lambda^2 / 2 )$. Note that in the $\pi K$ modes $| \lambda_u^s |$
is much smaller than $| \lambda_c^s |$. Taking
$V_{ub}=(3.82\pm0.15)\times 10^{-3}$ and
$V_{cb}=(41.79\pm0.63)\times 10^{-3}$ \cite{CKMfitter}, we have $| \lambda_u^s /
\lambda^s_c | = 0.021\pm0.001$.

The effective Hamiltonian for $\Delta S=0 (1)$ non-leptonic B
decays is given by
\beq \label{EH} H_{eff}=\frac{G_{F}}{\sqrt{2}} \sum_{q=u,c}
\lambda_{q}^{(s)}
                       \left(
                              C_{1} O^q_{1}+ C_{2} O^q_{2} + \sum_{i=3}^{10}
                              C_{i}O_{i}
                       \right),
\eeq
 where $O^{u(c)}_{1,2}$,  $O_{3,\dots, 6}$ and $O_{7,\dots
,10}$ are related to tree, QCD penguin and electro-weak penguin
sectors respectively and $C_i's$ are the corresponding short
distance Wilson coefficients.
In the SM, from the isospin structure of the effective
Hamiltonian, the ratio between electroweak penguin and tree
diagrams are fixed through \cite{NR}
\begin{eqnarray}
  R_{EW}=\frac{P_{EW} + P_{EW}^C}{T + C} & =  &  \frac{3}{2} \cdot
  \frac{C_9 + C_{10}}{C_1 + C_2} = -( 1.25 \pm 0.12 ) \times 10^{- 2} ,
  \label{pewSM}
\end{eqnarray}
for $\pi\pi$ modes. Where $T$, $C$, $P_{EW}$ and $P_{EW}^C$ are
diagrams with CKM matrix elements factorized out. $C_i$s stand for
the short distance Wilson coefficients at the scale of $\mu \simeq
m_b$. A direct consequence from this relation is that no direct CP
violation occurs in the $B\to \pi^-\pi^0$ decay, namely
 \beq
 & & A_{CP} (B\to \pi^-\pi^0) \simeq 0, \qquad \mbox{SM} \\
 & &  A_{CP} (B\to \pi^-\pi^0) \gg 0.1, \qquad \mbox{new physics}
 \eeq
as long as isospin symmetry holds at a few percent level. The latest
average data is $A_{CP} (B\to \pi^-\pi^0) =0.04\pm0.05$ which is not
precise enough to draw a robust conclusion. However, in the
factorization approach, it has been demonstrated that the SU(3)
symmetry breaking effects are small either because of the
cancellation between two combining factors of the decay constants and
form factors, namely $f_K f_0^{B\to \pi} \simeq f_{\pi} f_0^{B\to
K}$, or  the suppression by the heavy bottom meson mass
$(m_K^2-m_{\pi}^2)/m_B^2 \ll 1$\cite{NR}. The typical corrections
are less than $10\%$. Thus in flavor SU(3) limit, the relation of
eq.(\ref{pewSM}) should still hold in a good approximation in $\pi
K$ system, i.e., $R'_{EW} \simeq R_{EW}$\cite{SU3}.
This relation is free from hadronic
uncertainties and survives under elastic final state interactions
(FSIs) and inelastic FSIs through low isospin states such as $B\to
DD_{s}\to \pi\pi(K)$. It can directly confront the experiments
and allows us to explore new physics in hadronic charmless $B$
decays.

The current average data give the following results for the ratios
$R_c$ and $R_n$ in $ \pi K$ system:
 \beq
 R_c = 1.11\pm0.08 ,\quad
 R_n = 1.00\pm0.07,
 \eeq
which shows that $R_c$ and $R_n$ are all enhanced in comparison with
the previous values, the difference of two ratios is $\Delta R =
R_c-R_n = 0.11\pm 0.10$ which is diminished in comparison with the
previous result $\Delta R = 0.18\pm 0.10$. As shown in
ref. \cite{WZZ} that $\Delta R$ is dominated by $|P_{EW}'/P'|^2$,
a large deviation from the small value $\Delta R \simeq 0.02$
may indicate signal of new physics beyond SM. In general, the ratio
difference $\Delta R$ is a sensitive probe for new
physics. Note that the Belle collaboration reported almost the same $R_n$ and
$R_c$:
 \beq
R_c = 1.08\pm0.06\pm0.08 ,\quad
 R_n = 1.08\pm0.08^{+0.09}_{-0.08},\qquad \mbox{(Belle)}
 \eeq
the central values are consistent with the SM
estimation but the uncertainty is still large. Meanwhile
the BaBar collaboration reported the following results,
 \beq
R_c = 1.11\pm0.07\pm0.07 ,\quad
 R_n = 0.94\pm0.07\pm 0.05,\qquad \mbox{(BaBar)}
\eeq
From the present experiments, one can not yet rule out the
possibility of new physics.

\section{Enhanced Color-Suppressed Amplitudes from $B\to \pi\pi$}

We now discuss $B\to \pi\pi$ decays.  Using the diagrammatic
method, the CP-averaging branching ratios have the following
forms:
 \beq
Br(\pi^+\pi^-) &\simeq&
|\lambda_u|^2|T|^2+(|\lambda_u|^2+|\lambda_c|^2-2\cos{\gamma}|\lambda_u||\lambda_c|)|P|^2\non
   &&+2|\lambda_u||P||T|\cos{\delta_{T}}(|\lambda_c|\cos{\gamma}-|\lambda_u|),\non
 Br(\pi^0\pi^0) &\simeq& \frac{1}{2}
 [|\lambda_u|^2|C|^2+(|\lambda_u|^2
 +|\lambda_c|^2-2\cos{\gamma}|\lambda_u||\lambda_c|)|P-P_{EW}|^2]\non
 &&-2|\lambda_u||P-P_{EW}||C|\cos{\delta_{C}}(|\lambda_c|\cos{\gamma}-|\lambda_u|),\non
\frac{1}{\tau}Br(\pi^-\pi^0) &\simeq&
\frac{1}{2}|\lambda_u|^2|T+C|^2,
 \eeq
 where $\tau= \tau_B^-/\tau_B^0 = 1.086$ reflecting the
life-time difference. Here we have neglected the subleading diagrams
$P_{EW}^C$, $E$ and $P_A$ for simplicity.
$\delta_{C},\delta_{T}$ and $\delta_{EW}$ are the strong phases of $C$,
$T$ and $P_{EW}$ respectively.  The strong phase of $P$ is fixed to be
zero as an overall phase. The CP violation parameters $S$ and $C$ in
$B \to \pi^+\pi^-$ decays are introduced
 through the time-dependent decay rate difference:
 \beq
 A_{CP}(t) &=& \frac{\Gamma(\bar{B^0} \to \pi^+\pi^-)-\Gamma(B^0 \to
 \pi^+\pi^-)}{\Gamma(\bar{B^0} \to \pi^+\pi^-)+\Gamma(B^0 \to
 \pi^+\pi^-)}\non
 & \simeq & -a_{\epsilon} + (a_{\epsilon} + a_{\epsilon'})\cos{(\Delta m_B\cdot
 t)} + a_{\epsilon + \epsilon'} \sin{(\Delta m_B\cdot t)}, \non
 &\simeq & S\cdot\sin{(\Delta m_B\cdot t)} - C \cdot\cos{(\Delta m_B\cdot
 t)},\label{tdcpv}
 \eeq
 $\Delta m_B$ is the neutral B meson mass difference. The CP-violating
 quantities are defined as:
 \beq
 & & S = \frac{\im \lambda}{1+|\lambda|^2}= a_{\epsilon +
 \epsilon'},\quad \mbox{and} \quad
  C = \frac{1-|\lambda|^2}{1+|\lambda|^2} = -A_{CP} = -(a_{\epsilon} + a_{\epsilon'})
  \eeq
with $\lambda = e^{-2i\beta}(\bar{A}/A)$. Where the
rephase-invariant quantities $a_{\epsilon}$, $a_{\epsilon'}$ and
$a_{\epsilon + \epsilon'}$ \cite{PW} represent indirect, direct and
mixing-induced CP violations respectively. As $a_{\epsilon}\ll 1$
for neutral B system, we have $A_{CP} \simeq a_{\epsilon'}$ which
characterizes direct CP violation.

With the above equations, we can get the explicit expressions of
$A_{CP}(\pi^+\pi^-)\cdot Br(B \to \pi^+\pi^-)$ and $S_{\pi^+\pi^-}$ as:
 \beq
 && A_{CP}(B \to \pi^+ \pi^-)\cdot Br(B \to \pi^+\pi^-) \simeq
2|\lambda_u\lambda_c|\sin{\gamma}
  |T||P|\sin{\delta_{T}},\non
  & & S_{\pi^+\pi^-} \simeq
  \frac{2\kappa\cos{2(\beta+\gamma)}\cos{\delta_T}\sin{\gamma}-\sin{2(\beta+\gamma)}
  (1+2\kappa\sin{\delta_T}\sin{\gamma})}
  {1+2\kappa\sin{\delta_T}\sin{\gamma}
  +2\kappa^2\sin^2{\gamma}},
 \eeq
with  $\kappa = |\lambda_c P|/|\lambda_u T|$. Noticing the fact that $|P| \ll |T|$ and
$2|\lambda_u^d|(|\lambda_c^d| \cos{\gamma}-|\lambda_u^d|) \simeq
0.4 |\lambda_u^d|^2$ and considering the error of data, we can
safely ignore the cross term in branching ratio of $B \to
\pi^+\pi^-$. Similarly, we can also ignore the cross term in the
branching ratio of $\pi^0\pi^0$ and obtain in a good approximation
the following relations:
 \beq
 & & \frac{R_{-0}}{(1-R_{00})} \simeq \frac{1 +|C/T|^2 + 2
 |C/T|\cos{(\delta_{T}-\delta_{C})}}{1
 -|C/T|^2}
\eeq
with $R_{-0}\equiv 2Br(\pi^-\pi^0)/Br(\pi^+\pi^-)$.
Taking the experimental
data for the three branching ratios and considering the possible
range for $\cos{(\delta_{T}-\delta_{C})}\in [1,-1]$, we arrive at
the following constraint for the ratio $|C|/|T|$:
 \beq
  0.60 \leq \frac{|C|}{|T|}\leq 0.97 .
  \eeq

Using the three precise observed data points of $Br(\pi^+\pi^-),
A_{CP}(\pi^+\pi^-)$, $S_{\pi^+\pi^-}$ and taking the latest
experimental result for $\sin{(2\beta)}$ as an
input parameter\cite{HFAG}, we  get:
 \beq
|P| = 0.10\pm0.03,\quad |T| = 0.58\pm0.05, \quad \delta_T =
0.60\pm0.10
 \eeq
Noticing the positivity of the quantity:
 \beq
(|\lambda_u|^2+|\lambda_c|^2-2\cos{\gamma}|\lambda_u||\lambda_c|)|P|^2
+2|\lambda_u||P||T|\cos{\delta_{T}}(|\lambda_c|\cos{\gamma}-|\lambda_u|)
> 0 .
\eeq
 The above inequality holds for $|P/T|\geq 0.1$ which is true from the
 above analysis, i.e., $ Br(\pi^+\pi^-)/\tau_B > |\lambda_u|^2
 |T|^2$. We then yield a more strong constraint for the ratio: \beq
\frac{|C|}{|T|}\leq \sqrt {R_{0}} \equiv
\sqrt{\frac{2Br(\pi^0\pi^0)}{Br(\pi^+\pi^-)} }\simeq 0.70
 \eeq
Combining the above two constraints, we have
 \beq
 & & 0.6 \leq
|C|/|T| \leq 0.7
 \eeq
Note that the above numerical bounds are obtained by simply
taking the central values of the experimental data. When taking
into account the experimental errors, the allowed range could be
enlarged by $(10 \sim 20)\%$. The result is still larger than the
theoretical estimations $|C/T|\simeq0.1\sim0.2$ calculated from both
the QCD factorization approach\cite{QCDF} and perturbative QCD
approach\cite{PQCD}. Although the next to leading order
contributions calculated recently in QCD factorization show some
enhancement of $C$, it is still difficult to meet the current
data\cite{QCDFnlo}. Also a large color suppressed tree diagram is
independently favored by $\pi K$ and $K \eta^{(')}$ data
\cite{BH,WZ2,Keta}.

\section{Implications from new experimental results of $B \to K\pi$ decays}

The latest averaged data give $A_{CP}(\pi^+K^-) = -0.098\pm0.015,
A_{CP}(\pi^0\bar{K^0}) = -0.12\pm0.11$ and $A_{CP}(\pi^0K^-) =
0.05\pm0.03$. All these preliminary measurements are more
precise. However, there still exists significant differences between
two experiments. We shall make, basing on the new data, a
model-independent analysis to determine the hadronic amplitudes and
see whether there is any implication for new physics beyond the SM.

\begin{figure}[htbp]
\begin{center}
\begin{tabular}{cc}
\scalebox{1.0}{\epsfig{file=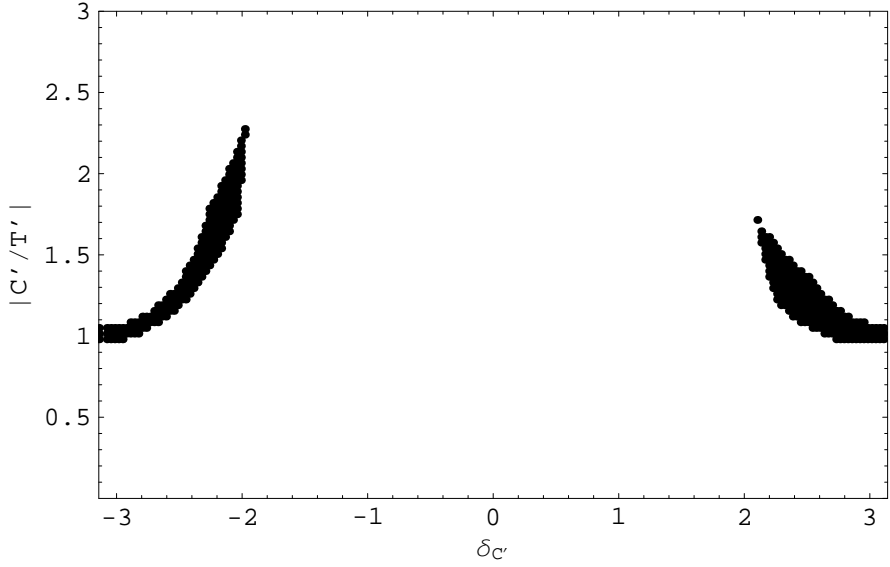}}
\end{tabular}
\caption{Allowed range for $|C'/T'|$ with $\delta_{C'}$ by using
five measured $\pi K$ data. } \label{Fig.1}
\end{center}
\end{figure}

In the $\pi K$ system, there are now five established experimental
observables, including four branching ratios and one direct CP
as of $B \to \pi^+ K^-$. Using the
diagrammatic language and neglecting the small contributions from
$P_{EW}^{C'},A',E'$, there are seven
free parameters, four magnitudes and three relative strong phases.
 Keeping isospin relation in Eq.(\ref{pewSM}) within the SM,
only five free parameters are left, namely three
magnitudes $|T'|,|C'|,|P'|$ and two relative strong phases
$\delta_{T'}$ and $\delta_{C'}$
, where we take the strong phase of $P'$ as an overall phase. In
this case, the data are enough to extract all these parameters. In
fact, by taking three data points of $Br(B\to \pi^-K^0), Br(B \to
K^+\pi^-)$ and $A_{CP}(K^+\pi^-)$, one can extract $|T'|, |P'|$
and $\delta_{T'}$. The numerical results are found to be
 \beq
 & & |T'| = 0.87\pm 0.18, \qquad \delta_{T'} =
0.33\pm 0.07,\qquad |P'| = 0.12\pm 0.02 .
 \eeq
 The other two data points are used to determine the color
 suppressed tree amplitude and it's strong phase.

In the first step, we shall work within SM. Neglecting the
color suppressed EW penguin,
and taking $Br(B \to \pi^0 \bar{K}^0)$ and $Br(B \to \pi^0 K^-)$
within $1\sigma$ error, we  find the allowed region for
$\delta_{C'}$ and $|C'/T'|$. The results are plotted in Fig.1.
\begin{figure}[htbp]
\begin{center}
\begin{tabular}{cc}
\scalebox{1.0}{\epsfig{file=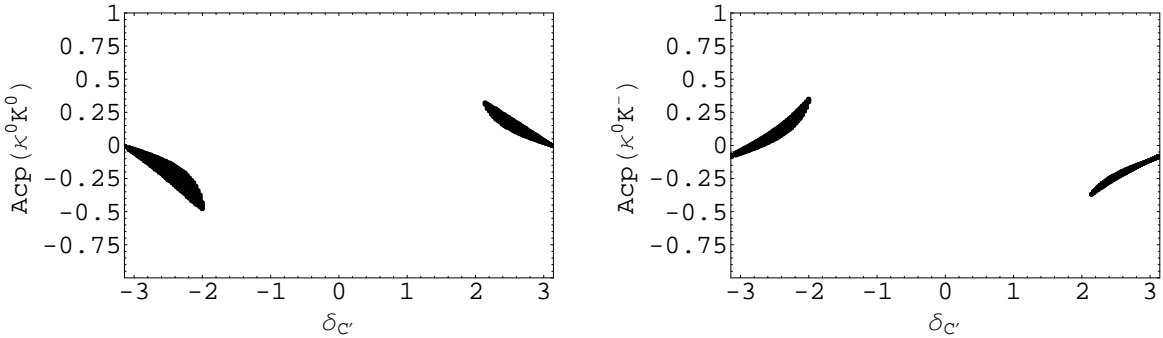}}
\end{tabular}
\caption{Allowed range for $A_{cp}(\pi^0\bar{K}^0)$ and
$A_{cp}(\pi^0K^-)$ as function of $\delta_{C'}$ by  using five
measured $\pi K$ data.} \label{Fig.2}
\end{center}
\end{figure}
In obtaining the figure,
we let $\delta_C'$ vary  in the range $[-\pi,\pi]$ and $|C'/T'|$
in  $[0,10]$. The result indicates that a large strong phase of
the color suppressed tree diagram is necessary to explain the
experiments and there exists two allowed regions with opposite
signs of $\delta_{C'}$ but similar size of $|C'|$ around $2.0\sim
3.0$. As for the ratio $|C'/T'|$, the minimal value is about
$|C'/T'|\simeq 1$ with $\delta_{C'} \approx \pm\pi$, the whole
allowed range is from $1.0 $ to $2.4$. So the large $C'$ puzzle is
still there though the minimal value can be reduced to about
unity. When only taking the latest data from the Belle
collaboration, the ratio $|C'/T'|$ can be further reduced and the
minimal size can reach $|C'/T'| \simeq 0.74$ which is still large.

The CP asymmetry in $\pi K$ decays can be expressed as follows
 \beq
 A_{CP}(B \to \pi^+ K^-) & \cdot & Br(B \to \pi^+ K^-) \non
   &\simeq&  -2|\lambda_u^s\lambda_c^s|\sin{\gamma} |T'||P'|\sin{\delta_{T'}}
 ,\non
A_{CP}(B \to \pi^0 \bar{K^0}) & \cdot & Br(B \to \pi^0 \bar{K^0})
\non
  &\simeq&  |\lambda_u^s\lambda_c^s|\sin{\gamma}|C'|
[|P'|\sin{\delta_{C'}}+|P_{EW}'|\sin{(\delta_{C'}-\delta_{EW'})}],\\
\frac{1}{\tau}A_{CP}(B \to \pi^0 K^-)& \cdot & Br(B \to \pi^0 K^-)
                      \simeq
                      -|\lambda_u^s\lambda_c^s|\sin{\gamma}\non
                      & \cdot &
                       [ |T'| (|P'|\sin{\delta_{T'}}
                        -|P_{EW}'|\sin{(\delta_{T'}-\delta_{EW'})}
                        ) \non
                        && + |C'| (|P'|\sin{\delta_{C'}} -|P_{EW}'|\sin{(\delta_{C'}-\delta_{EW'})}
                        )]\nonumber
 \eeq
 The expression of mixing-induced CP-violating parameter $S_{\pi K_S}$ is
 \beq
 S_{\pi^0K_S} &\simeq& \sin{(2\beta)} +
 2  r_C'  \cos{(2\beta)}\cos{\delta_{C'}}\sin{\gamma}-
 2  r_C'^2 \sin{(2\beta)}\sin^2{\gamma}\non
 &&- r_C'^2 \cos{(2\beta)}\cos{(2\delta_{C'})}\sin{(2\gamma)}  - 2
  r_C' r_{EW}' \cos{(2\beta)}\cos{(\delta_{C'}+\delta_{EW'})}\sin{\gamma},
 \eeq
where $r_C'\simeq |{\lambda_u^s}/{\lambda_c^s}||C'/P'|$ and
$r_{EW}' = |P_{EW}'/P'|$.  The corresponding predictions for
$A_{CP}(\pi^0\bar{K}^0)$ and $A_{CP}(\pi^0 K^-)$ are given in
Fig.2. It shows that there are two solutions corresponding to the
sign of $\delta_{C'}$
 \beq
\mbox{for } \delta_{C'} < 0: \non & &-0.08 < A_{CP}(\pi^0 K^-) <
0.39,\non
 & &-0.50 < A_{CP}(\pi^0\bar{K}^0) < 0, \non
\mbox{for }\delta_{C'} > 0: \non
 & &
-0.34 < A_{CP}(\pi^0 K^-) < -0.10,\non
 & &0 < A_{CP}(\pi^0\bar{K}^0) <0.35,
 \eeq
where $A_{CP}(\pi^0\bar{K}^0)$ and $ A_{CP}(\pi^0 K^-)$ almost
have opposite signs. In Fig.3, the mixing induced CP asymmetry
$S_{\pi K_S}$ as function of strong phase $\delta_{C'}$ is given.
One finds that for both positive or negative $\delta_{C'}$, the
resulting mixing CP violation $S_{\pi K_S}$ is the same, because
it depends only on $\cos{\delta_{C'}}$
 \beq
S_{\pi^0K_S} = 0.55\pm0.07
 \eeq
\begin{figure}[htbp]
\begin{center}
\begin{tabular}{cc}
\scalebox{1.0}{\epsfig{file=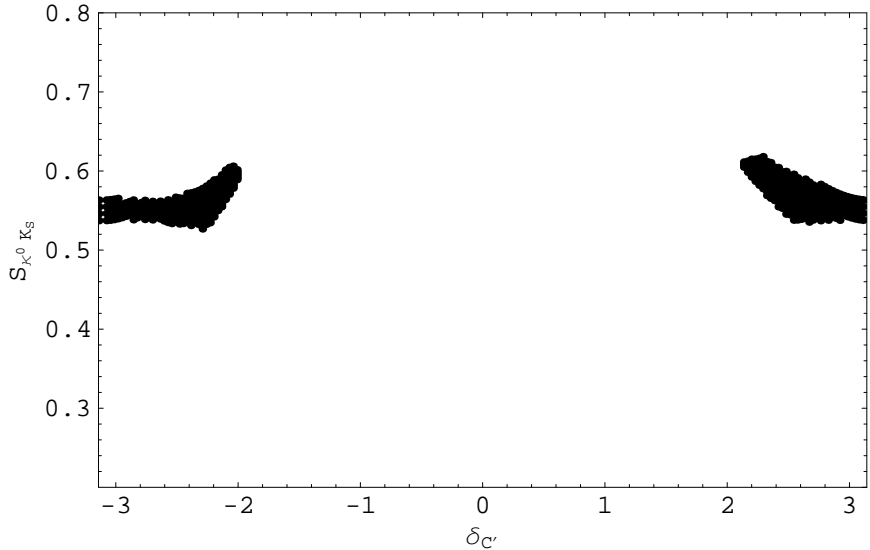}}
\end{tabular}
\caption{The allowed range for mixing induced CP asymmetry
$S_{\pi^0 K_S}$ as a function of strong phase $\delta_{C'}$ by
using five measured $\pi K$ data.} \label{Fig.3}
\end{center}
\end{figure}
From the above discussions, we see that from the measured five data,
there is still significant uncertainties in determining the
magnitude of $|C'/T'|$ and predicting for the direct CP violation.
In order to tighten the constraints, we try to add another data
point of $A_{CP}(\pi^0 K^-)$. Note that the preliminary result show
that the Babar and Belle's results are consistent with each other,
$A_{CP}(\pi^0 K^-) = 0.016\pm 0.041\pm 0.010(\mbox{Babar})$ and
$A_{CP}(\pi^0 K^-) = 0.07\pm 0.03\pm 0.01(\mbox{Belle})$. With this
extra data point included, a very strong constraint on $|C'/T'|$ is
found.  The allowed region for $\delta_{C'}$ and $|C'/T'|$ are given
in Fig.4, where we scan all the possible solutions to meet six data
points within the region of
$\delta_{C'}\in [-\pi,\pi]$ and $|C'/T'|\in[0,10]$. 
average data within $1\sigma$ error.  The figure shows that
$|C'/T'|$ and $\delta_{C'}$ can be well determined with $|C'/T'| =
1.16\pm 0.08, \delta_{C'} = -2.65\pm 0.10$. The positive
$\delta_{C'}$ solution is excluded completely, and there is no
two-fold ambiguity in the prediction of CP asymmetry for
$\pi^0\bar{K}^0$
 \beq
 A_{CP}(\pi^0\bar{K}^0) &=& -0.15\pm0.03,\non
 S_{\pi^0K_S} &=& 0.55\pm0.03 .
 \eeq

\begin{figure}[htbp]
\begin{center}
\begin{tabular}{ll}
\scalebox{1.0}{\epsfig{file=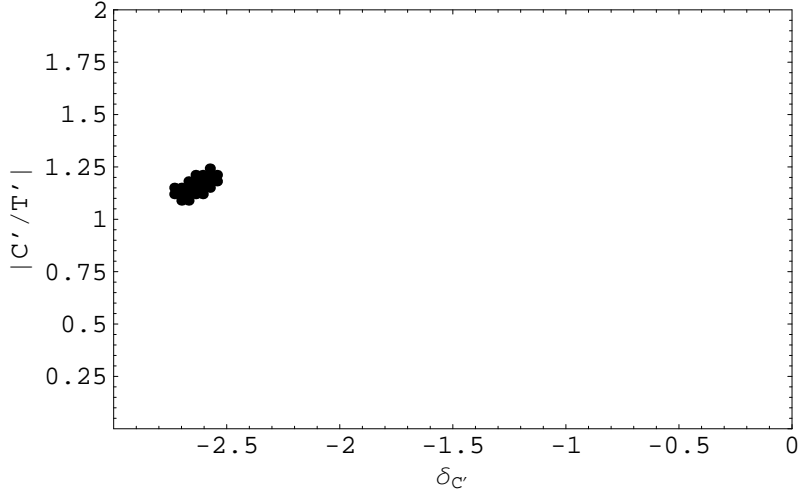}}
\end{tabular}
\caption{The allowed range for $|C'/T'|$ and $\delta_{C'}$ by
using five measured $\pi K$ data and the preliminary data of
$A_{cp}(\pi^0K^-)$} \label{Fig.4}
\end{center}
\end{figure}

The prediction coincides with the preliminary data at $1\sigma$
error and also  the results obtained by using the sum rules in $\pi
K$ system\cite{sumrule}. Although $|C'/T'|$ is moving towards the SM
value, a large value around unity with a large negative strong phase
$\delta_{C'} \simeq -2.65$ is still inevitable. Obviously, the
obtained ratio $|C'/T'| \simeq 1.2 $ is much larger than the
theoretical estimation using QCD factorization or pQCD method
\cite{QCDF,PQCD,QCDFnlo}. It is  almost twice as large as $|C/T|$
extracted from $\pi\pi$ system, which may indicate a breakdown of
flavor SU(3) symmetry. Unlike in the pQCD and QCD factorization
calculations, the soft collinear effective theory (SCET) shows that
the color suppressed amplitudes can be of similar size to the tree
amplitudes in \cite{scet1,scet2}, which may provide a dynamic QCD
explanation for large $C \sim T$, but it depends on two unknown
parameters $\zeta_J$ and $\zeta$, and the predicted strong phase is
smaller than the one obtained directly from the data, as shown in
ref.\cite{LM}, it also overshoots the bound of the $B\to
\rho^0\rho^0$ branch ratio and deteriorates the predictions for
$B\to \pi K$ direct CP violations .

\section{New physics effects}

In this section, we consider two kinds of phase effects in the
presence of new physics, i.e., there is a new CP phase or a new
strong phase in electroweak penguin sector, but the magnitude of
$P_{EW}'$ remains to be unchanged $|P_{EW}'| = |R_{EW}'||T'+C'|$.
For the case with an enhanced electroweak penguin amplitude has
widely been discussed, we shall not discuss its effects here, it is
referred to the recent papers in refs.\cite{BF,MY,WZ1,WZ2,WZZ}.
For the scenario with  a new CP violating phase, we add $\phi_{NP}$
to $P_{EW}'$, i.e., $P_{EW}' =
|P_{EW}'|e^{i(\delta_{EW'}+\phi_{NP})}$, the corresponding
expressions of branching ratios and CP asymmetries are changed
accordingly.  For example, in $\pi^0 \bar{K^0}$ mode:
 \beq
A_{CP}(B \to \pi^0 \bar{K^0}) & \cdot & Br(B \to \pi^0 \bar{K^0})
\non
 &\simeq&|\lambda_u^s\lambda_c^s|[\sin{\gamma}|C'P'|\sin{\delta_{C'}}
+\sin{(\gamma+\phi_{NP})}|C'P_{EW}'|\sin{(\delta_{C'}-\delta_{EW'})}]\non
 &&-[|\lambda_u^s|^2 + |\lambda_c^s|^2+
 2|\lambda_u^s\lambda_c^s|]\sin{\phi_{NP}}\sin{\delta_{EW'}}|P'P_{EW}'| .
 \eeq
We use the whole six data points and scan all the allowed values
of $\phi_{NP}$ and $\delta_{C'}$ which meet the data within
$1\sigma$ error. The allowed region of $|C'/T'|$ is found to be
$[0.40,2.40]$ with $\delta_{C'} = -1.5\pm0.7 $. For a large weak
phase $\phi_{NP} \approx \pm(2.6\pm0.4)$ , the ratio $|C'/T'|$ can
be strongly reduced to the range $0.40\sim 0.80$ which is similar
to the result of $|C/T|$ obtained in the $\pi\pi$ system. The
corresponding CP asymmetries in this case are:
 \beq
 A_{CP}(\pi^0\bar{K}^0) &=& -0.22\pm0.12,\non
 S_{\pi^0K_S} &=& 0.60\pm0.20 .
 \eeq
In ref. \cite{BFRS}, the authors introduced such a new CP phase as a
new physics scenario in electroweak penguin sector, and found that
$\phi_{NP} \simeq \pm \pi/2$ is needed to meet the $B \to \pi K$
data. Here we adopt the latest experimental data and find that a
much larger $\phi_{NP} \approx \pm(2.6\pm0.4)$ is required to give a
consistent explanation for the data when keeping SU(3) symmetry.

In the other scenario, we add a new strong phase $\delta_{NP'}$ to
$P_{EW}'$, namely the isospin relation will be broken by an extra
phase factor, $P_{EW}' = R_{EW}'(T'+C')e^{i\delta_{NP'}}$. By
redefining it as $P_{EW}' = |R_{EW}'(T'+C')|e^{i\delta_{EW'}}$,
similar results can be obtained as introducing a new CP phase in
$P_{EW}'$. It is interesting to find that a new strong phase
$\delta_{EW'} = -2.70\pm 0.30$ can also reduce $|C'/T'|$ to
$0.40\sim0.80$. As a consequence, the CP asymmetries are predicted
to be \beq A_{CP}(\pi^0\bar{K}^0) &=& -0.10\pm0.10,\non
S_{\pi^0K_S} &=& 0.57\pm0.12 .\eeq
Thus we find that either a new weak phase or a new isospin relation
broken strong phase in electroweak penguin diagrams can
significantly reduce the ratio $|C'/T'|$ to be close to the ratio
$|C/T|$ in $\pi\pi$ system and meet all the present experimental
data well within $1\sigma$ error. While $|C'/T'| \simeq 0.40\sim
0.80$ is still $2\sim 3$ times larger than the theoretical
evaluation calculated by using QCD factorization and PQCD methods,
also a large weak phase or strong phase around $\pm(2\sim3)$ is
necessary. Nevertheless, the predictions for the direct CP asymmetry
of $B \to \pi^0\bar{K}^0$ is consistent with the averaged data
though BaBar and Belle did not give very consistent values and the
errors are still large $A_{CP}(\pi^0\bar{K}^0) =
-0.20\pm0.16\pm0.03$(BaBar) and $A_{CP}(\pi^0\bar{K}^0) =
-0.05\pm0.14\pm0.05$(Belle)\cite{ichep06}. As for $S_{\pi K_S}$, in
both cases, a large value around $0.6$ is obtained in comparison
with the experimental result $0.33\pm0.21$, they are consistent
within $1.5\sigma$ error. It is noted that both cases can bring out
similar effects: a) reducing the ratio $|C'/T'|$ to a reasonable
value where SU(3) symmetry holds; b) leading to almost the same
prediction for $S_{\pi^0 K_S}\approx 0.60$ and a consistent
prediction for $A_{CP}(\pi^0\bar{K}^0)$. However, adding a new
strong phase in $P_{EW}$ will lead to a nonzero $A_{CP}(\pi^-\pi^0)$
at a few percent level, which can be used  to distinguish the two
type of scenarios in the future.

\section{Conclusions}

In summary, we have presented a model-independent analytical
analysis for $B \to \pi\pi$ and $\pi K$ decays based on the latest
data. We obtained the ratio between color suppressed tree diagram
and tree diagram $0.6\leq|C/T|\leq0.7$ in $\pi\pi$ system through
model-independent analysis and found that the latest data makes it
move closer to the SM estimation, but was still large. We made an
similar analysis in $\pi K$ system and found $|C'/T'|\approx 1.16$
in comparison with the previous value $|C'/T'|\approx 2.0 $ and a
large negative strong phase $\delta_{C'} \simeq -2.65$. The
predictions for CP asymmetries in $\pi^0 K^0$ are
$A_{CP}(\pi^0\bar{K}^0) = -0.15\pm0.03$ and $S_{\pi^0K_S} =
0.55\pm0.03$, which are consistent with the present data within
$1\sigma$ and $1.5\sigma$ respectively. We also considered two kinds
of new physics scenarios with a new CP phase and a new strong phase
in electroweak penguin sector.  In our present analysis with the
latest experiment data, we found that with a new CP phase $\phi_{NP}
= \pm(2.2\sim 3.0)$, the ratio $|C'/T'|$ could be  reduced to $0.40
\sim 0.80$ that was similar to that in $\pi\pi$ system, and the
strong phase $\delta_{C'}$ could also be reduced to about $-1.5$.
Alternatively , an extra strong phase in electroweak penguin sector
which breaks the isospin symmetry in phase could also reduce the
ratio $|C'/T'|$ to about $0.40\sim0.80$ and coincided with the data
well. Thus, new physics may help to explain the discrepancy between
$|C/T|$ and $|C'/T'|$ in $\pi\pi$ and $\pi K$ systems. Recently, it
has been shown that new physics effects in $P'_{EW}$ can be
effectively reparameterized into $C'$\cite{sinha}, thus
 a large $|C'/T'|$ may also be a consequence of new physics.
The two kinds of scenarios lead to a consistent prediction for
$A_{CP}(\pi^0 \bar{K}^0)$ and similar for $S_{\pi K_S} \simeq 0.60$
which is consistent with the data within $1.5\sigma$ error.
A more precise measurement of $A_{CP}(\pi^-\pi^0)$ may not only help
us to signal out new physics but also distinguish these two kinds of new
 scenarios in the near future.

\acknowledgments

\label{ACK}

This work is supported in part by the National Science Foundation
of China (NSFC) under the grant 10475105, 10491306, and the
Project of Knowledge Innovation Program (PKIP) of Chinese Academy
of Sciences. YFZ is supported by JSPS foundation.

\end{document}